# Minimal Specifications for Non-Human Primate MRI: Challenges in Standardizing and Harmonizing Data Collection

## Authors
Joonas A. Autio[1], Qi Zhu[2,3], Xiaolian Li[2], Matthew F. Glasser[4,5], Caspar M. Schwiedrzik[6,7], Damien A. Fair[8], Jan Zimmermann[8], Essa Yacoub[8], Ravi S. Menon[9], David C. Van Essen[5], Takuya Hayashi[1], Brian Russ[10,11,12,*], Wim Vanduffel[2,13,14,15,*]

## Affiliations
[1]Laboratory for Brain Connectomics Imaging, RIKEN Center for Biosystems Dynamics Research, Kobe, Japan
[2]Laboratory for Neuro- and Psychophysiology, Department of Neurosciences, KU Leuven Medical School, Leuven 3000, Belgium
[3]Cognitive Neuroimaging Unit, INSERM, CEA, Université Paris-Saclay, NeuroSpin Center, 91191 Gif/Yvette, France
[4]Departments of Radiology and [5]Neuroscience, Washington University School of Medicine, St. Louis, MO, USA
[6]Neural Circuits and Cognition Lab, European Neuroscience Institute Göttingen – A Joint Initiative of the University Medical Center Göttingen and the Max Planck Society, Grisebachstraße 5, 37077 Göttingen, Germany
[7]Perception and Plasticity Group, German Primate Center – Leibniz Institute for Primate Research, Kellnerweg 4, 37077 Göttingen, Germany
[8]Center for Magnetic Resonance Research, Department of Radiology, University of Minnesota, Minneapolis, MN, USA
[9]Centre for Functional and Metabolic Mapping, Western University, London, ON, Canada
[10]Department of Psychiatry, New York University Lagone, New York City, New York, USA
[11]Center for the Biomedical Imaging and Neuromodulation, Nathan Kline Institute, Orangeburg, New York, USA
[12]Department of Neuroscience, Icahn School of Medicine, Mount Sinai, New York City, New York, USA
[13]Leuven Brain Institute, KU Leuven, Leuven 3000, Belgium
[14]Athinoula A. Martinos Center for Biomedical Imaging, Massachusetts General Hospital, Charlestown, MA 02129, USA
[15]Department of Radiology, Harvard Medical School, Boston, MA 02144, USA

**\*Equal contributions**

## Corresponding author
Joonas A. Autio
Laboratory for Brain Connectomics Imaging
RIKEN Center for Biosystems Dynamics Research
6-7-3 MI R&D Center 3F, Minatojima-minamimachi
Chuo-ku, Kobe 650-0047, Japan

**Abstract**

Recent methodological advances in MRI have enabled substantial growth in neuroimaging studies of non-human primates (NHPs), while open data-sharing through the PRIME-DE initiative has increased the availability of NHP MRI data and the need for robust multi-subject multi-center analyses. Streamlined acquisition and analysis protocols would accelerate and improve these efforts. However, consensus on minimal standards for data acquisition protocols and analysis pipelines for NHP imaging remains to be established, particularly for multi-center studies. Here, we draw parallels between NHP and human neuroimaging and provide minimal guidelines for harmonizing and standardizing data acquisition. We advocate robust translation of widely used open-access toolkits that are well established for analyzing human data. We also encourage the use of validated, automated pre-processing tools for analyzing NHP data sets. These guidelines aim to refine methodological and analytical strategies for small and large-scale NHP neuroimaging data. This will improve reproducibility of results, and accelerate the convergence between NHP and human neuroimaging strategies which will ultimately benefit fundamental and translational brain science.

**Highlights**

- Non-human primate MRI standardization.
- Poor reproducibility in non-human primate resting-state fMRI.
- Standards enable improved and more reproducible MRI measures.
- Convergence between non-human primate and human neuroimaging strategies.

## 1. Introduction

Non-human primate (NHP) magnetic resonance imaging (MRI) has come a long way in the last two decades from being a nascent field (Disbrow et al., 2000; Logothetis et al., 1999; Stefanacci et al., 1998; Vanduffel et al., 2001) to a growing and maturing research field in many institutes across the globe (Milham et al., 2020). While methods have been shared and improved, most research groups rely on custom-designed experimental set-ups and protocols. Such customization is not surprising in a newly emerging research domain, but it hampers comparison of results across studies, which can lead to suboptimal paradigms and a waste of resources. Concurrently, imaging of human subjects increased dramatically, in part through the harmonization of acquisition and data standards. In particular, initiatives like the WU-Minn Human Connectome Project (HCP) (Van Essen et al., 2013), UK BioBank (Miller et al., 2016) and the Adolescent Brain Cognitive Development (ABCD) study (Casey et al., 2018; Hagler et al., 2019) have helped to implement large-scale collaborative projects that focus on collecting standardized data sets across imaging sites. These initiatives established a set of minimal data standards and practices (Glasser et al., 2016b; Smith et al., 2013; Uğurbil et al., 2013; D. C. Van Essen et al., 2012), which has facilitated human imaging research around the world. We believe it is timely to translate the lessons from these human imaging initiatives to the monkey imaging community. This will boost the growing NHP imaging field by sharing standardized data across multiple sites and studies and allowing meta-analyzes on data that cannot be acquired in any individual NHP laboratory.

Despite remarkable advances in MRI over the last two decades, there are still inherent technical, physiological, behavioral and analytical challenges in standardizing NHP neuroimaging (Vanduffel et al., 2014). While there has been a convergence to just a few vendor hardware and software platforms for human studies over the past decade, in part due to initiatives such as the HCP and ABCD studies, approaches for NHP imaging are far less standardized, in part due to the use of pre-clinical hardware and software that requires customization. The objective of this manuscript is to promote the type of convergence that has occurred in human neuroimaging.

The first step towards standardized data acquisition across primate species is to recognize that the spatial resolution should be adjusted according to the neuroanatomical size of the brain. Brain volume is approximately 1,300 $cm^3$ in humans, 330 $cm^3$ in chimpanzees, 170 $cm^3$ in baboons, 100 $cm^3$ in macaques, 7.5 $cm^3$ in common marmosets and 1.7 $cm^3$ in the mouse lemur, the smallest experimentally-used primate. The number of neurons also ranges over more than two log units across these primates (i.e. human and mouse lemur have 86 and 0.2 billion neurons, respectively) (Herculano-Houzel, 2009). The huge difference in brain volume across primates represents a major technical challenge for standardization of data acquisition protocols as it drives a need for species-specific data acquisition hardware. The attractiveness of using monkeys for accelerated studies of development also imposes constraints on data-acquisition hardware as the brain and body sizes can change substantially over relatively short periods of time. Moreover, the MRI measurements are affected by different hardware choices (e.g., bore size, magnetic field strength, gradient field specifications, design of radiofrequency (RF) transmit coil and number of RF receive channels), acquisition parameters (e.g., voxel size, repetition time, echo-time, multiband acceleration, in-plane acceleration and b-value) and vendor-specific image reconstruction algorithms (e.g., un-aliasing algorithms, phase and frequency corrections, coil combinations, image filters and receiver biasfield corrections). In addition, magnetic field ($B_0$) strength strongly influences tissue contrast generation mechanisms in structural (e.g. dipolar relaxation) (Marques et al., 2017; Wang et al., 2020) and functional images which influences tissue segmentation (Lüsebrink et al., 2013) and sensitivity to neural activity (Yacoub et al., 2020), respectively. Finally, unlike human functional MRI (fMRI) experiments, NHPs are often scanned while anesthetized with a variety of agents, which strongly impacts brain activity (e.g., motor control, sensory processing, cognitive control), physiology, neurovascular coupling, and functional connectivity (FC) (Paasonen et al., 2018; Xu et al., 2018), emphasizing the importance of functional imaging in awake behaving monkeys (Belcher et al., 2013; Kagan et al., 2010; Landi and Freiwald, 2017; Logothetis et al., 1999; Miyamoto et al., 2017; Park et al., 2017; Schaeffer et al., 2019; Silva et al., 2011; Vanduffel et al., 2001; Wilke et al., 2012).

Another major challenge to harmonizing NHP studies is the lack of standardized image pre-processing software. Currently, research groups commonly use a mixture of tools borrowed from SPM, FSL, AFNI, etc. Thus, custom image-processing pipelines and different statistical analysis routines are often used in studies of NHPs, which can have substantial effects on results and reduce study reproducibility (Botvinik-Nezer et al., 2020; Carp, 2012; Eklund et al., 2016). The need for custom-made pre-processing pipelines emerged, in part, from lack of standardized geometric models for brain tissues (i.e. cerebral and cerebellar cortices and hippocampal complex) that are critical to account for the 3D geometry of highly folded cortices (Fischl and Sereno, 2018; Van Essen, 2002), to account for inter-subject variability (Coalson et al., 2018; Robinson et al., 2018), and to make use of CIFTI grayordinates that represent cerebral cortex by surface vertices and subcortical gray matter by voxels (Glasser et al., 2013). Moreover, laboratory-specific non-biological measurement biases need to be removed from the data to improve detection of biologically important features across imaging centers and species using retrospective harmonization strategies. So far, only limited attempts have been made to harmonize data acquisition and analysis across preclinical imaging centers (Immonen et al., 2019) and species (Autio et al., 2020; Warrington et al., 2020) and the development of translational informatics and information technology platforms for NHPs remains important to improve translational science (Milham et al., 2018; Van Essen et al., 2012; Van Essen et al., 2019).

Here, we specify NHP imaging recommendations based on the current state of NHP MRI. Recent advances in preclinical scanners and improved availability of multi-channel receive coils have made it possible to capitalize on accelerated imaging in NHPs (Ekstrom et al., 2008). Moreover, as the availability of NHP imaging data rapidly increases via open data-sharing platforms, including the PRIMatE Data Exchange (PRIME-DE) (Milham et al., 2018), there is an increased necessity for validated, automated and robust multi-subject analyses. Robust translation of human oriented open-access analytical solutions to NHP brain imaging, however, requires significant adaptations of image acquisition and analytical standards. By providing minimal image acquisition standards the NHP neuroimaging community can refine their methodological and analytical strategies to improve robustness and reproducibility. During this process, we hope to accelerate the convergence between NHP and human neuroimaging communities and to progress towards an era of improved comparative brain science, with important ramifications for both fundamental and translational research.

## 2. Suggestions for minimum acquisition requirements

The minimum specifications for NHP MRI (Box 1) include scaling resolution (anatomical and functional) to respect the thickness of cerebral cortex, transmit ($B_1+$) and receive ($B_1-$) RF biasfield corrections to ensure uniformity of contrast and intensity across the field of view, echo-planar imaging (EPI) acquisition in both phase encoding (PE) directions and acquisition of a $B_0$ field-map. Phased-array coils are a prerequisite for accelerated imaging and have been successfully applied in studies of NHPs (Autio et al., 2020; Ekstrom et al., 2008; Freiwald and Tsao, 2010; Gao et al., 2020; Gilbert et al., 2016; Janssens et al., 2013, 2012; Yacoub et al., 2020) (Box 2). Auxiliary minimum specifications include physiological monitoring in studies of anesthetized NHPs and behavioral monitoring and control in awake NHPs, which have important implications for MR data quality. The rationale and specifics of each minimum criterion are articulated below. Importantly, since most of these minimum criteria are utilized and tested by the majority of recent large-scale human MRI consortia (i.e. HCP, UK BioBank, ABCD, Brain/MINDS), application of these criteria in NHPs should improve reproducibility and prospects for translational science.

**Box 1 | Minimum specifications for non-human primate MRI**

MRI provides an invaluable tool to decipher the macro- and mesoscopic whole-brain anatomy and multi-synaptic network activities thought to underlie species-specific behavior. However, MRI has numerous biases and artefacts that need to be corrected for robust image alignment across modalities (i.e. functional to anatomical) and subjects. Challenges in NHPs arise partially from prominent differences in MR hardware specifications and data acquisition protocols across imaging sites. To improve robustness and reproducibility, the PRIME-DE consortium recommends the following set of minimal specifications:

**Animal preparation**
- Ethical animal protocol review by an independent review board (e.g., NIH or ILAR guidelines).
- Awake: intensive mock scanner training (e.g. operant behavior contingent on eye and hand positions); animal acclimatization inside the scanner (e.g. pilot scan sessions mainly to improve the animal's behavior); continuous (ideally automated) monitoring using MR compatible cameras for any signs of discomfort and general cognitive state.
- Anesthetized: Monitoring subject physiology (respiration, heart-rate, $EtCO_2$, pulse oximetry, blood pressure and body temperature) and cognitive state.

**Data acquisition**
- 3T field strength or higher
- Multichannel RF receiver coil (for dMRI & fMRI, ideally with multi-band pulse sequences); adhere Specific Absorption Rate (SAR) safety regulations (level 1); gradients 60 mT/m (preferably more for small NHPs).
- Fine-tuning of the $B_0$ shims (manual iteration or with algorithms scaled to the brain size).
- Prescan normalized images for $B_1$- intensity bias correction.
- Anatomical isotropic resolution corresponding to at most half of the minimum cortical thickness (e.g. 0.5 mm in macaque); 3D whole-brain coverage; T1w MPRAGE and T2w SPACE (matched resolution) (**Fig. 1**).
- A $B_1$+ measure when using surface RF transmitters.
- $B_0$ field-map; spin-echo EPI (preferred) or traditional gradient-echo; field map scans exactly matched with geometry and distortion properties of the fMRI scans (**Fig. 3a,b,c**).
- FMRI isotropic resolution at less than median cortical thickness of 2.1 mm (ideally 1.25 mm or less if achievable) (**Fig. 4a**); both phase encoding directions (in the orientation that allows minimized echo spacing); Ernst angle; improve CNR using contrast-enhanced cerebral blood volume weighted fMRI with minimized echo time (**Fig. 9**); resting-state data 30 min or more (**Fig. 5, 6**); minimized TR (ideally ~1 s or less if achievable for improved denoising and statistics).

**Data analysis**
- Pre-processing: automatized pipelines; $B_1$ biasfield correction (**Fig. 1a,b,c**); $B_0$ geometric correction (**Fig. 3c,d**); validate geometric correction (e.g. FSL TopUp configuration (**Fig. 3f**)); frame-to-frame distortion correction for awake fMRI; brain masking; segmentation; cortical surface generation (**Fig. 1d**); motion correction; cross-modal registration; avoid spatial smoothing in the volume space; spatial and temporal artefact removal (**Fig. 7a,c,d**); quality assurance (**Fig. 7f**).
- Post-processing: statistical toolboxes firmly established for human neuroimaging; multiple comparison correction (for best statistical practices, see (Eklund et al., 2016; Nichols et al., 2017)).

All animal experiments reviewed herein were conducted in accordance with the institutional guidelines for animal experiments and animals were maintained and handled in accordance with the Guide for the Care and Use of Laboratory Animals of the Institute of Laboratory Animal Resources (ILAR; Washington, DC, USA). Data from the KU Leuven group were acquired in agreement with institutional (KU Leuven Medical School: Ethische Commissie Dierproeven), national and European guidelines (Directive 2010/63/EU). We also used HCP data as a reference for data quality. The use of HCP data was approved by the institutional ethical committee (KOBE-IRB-16-24).

2.1. Structural MRI

Structural images provide a fundamental basis not only for anatomical measures but also for functional and diffusion measures and multi-subject statistics, yet the minimal structural image acquisition criteria are not well established. First, spatial resolution (isotropic) should be at least half of the minimum cortical thickness to enable robust segmentation, while higher resolutions further improve the discrimination of thin regions of cerebral cortex or regions with thin gyral blades of white matter (Glasser et al., 2013). In macaque monkeys, for example, the thinnest part of the cerebral cortex is approximately 1.0 mm, we therefore recommend a minimum structural (isotropic) image resolution of 0.5 mm. Second, 3D whole-brain coverage is important to improve inter-subject registration and to boost SNR (relative to 2D acquisitions). Third, acquiring both MPRAGE T1w and SPACE T2w images is recommended to discriminate among tissue types because they provide better CNR in white matter and fluid-filled regions, respectively (Fig. 1a). Moreover, T2w is advantageous for distinguishing dura and blood vessels. Fourth, imaging parameters that influence the contrast among white matter, grey matter and cerebrospinal fluid (e.g. inversion time, flip-angle, echo time) should be adjusted to yield sufficient contrast-to-noise ratio (CNR) (i.e. practically determined by robust automatic segmentation). We recommend to optimize MPRAGE inversion time (~900 ms in macaque at 3T; see Supp. Fig. 1) and SPACE echo-time (~500 ms in macaque at 3T) to improve segmentation and intracortical myelin contrast. For both sequences, we recommend matching spatial resolutions and using fat insensitive RF excitation to improve tissue classification in FreeSurfer (Glasser et al., 2013). Depending on the $B_0$ strength, head coil, and system performance, it is advised to acquire multiple T1w (between 3-10) and T2w (between 1-4) images for increasing SNR, ideally during a dedicated anatomical session in which the animal is anesthetized. We recommend that NHP researchers dedicate one, or more sessions, to specifically acquire a high-resolution set of anatomical images that can be used for both anatomical analyses (i.e. morphology, surface creation, myelin mapping) and registration purposes.

The $B_1+$ field homogeneity is an important factor to ensure consistent contrast and signal intensity between grey matter, white matter, and CSF in T1w and T2w images, because these factors have an influence upon tissue segmentation (Donahue et al., 2018; Fischl, 2012; Glasser et al., 2013). Homogeneous $B_1+$ is also important for myelin mapping, calculated from T1w and T2w ratios, which is valuable for distinguishing anatomical landmarks (e.g., area MT, auditory and primary somatosensory cortex) (Glasser et al., 2016a; Glasser and Van Essen, 2011), and even smaller sub-compartments such as inter-stripes in area V2 (Li et al., 2019). This is a major technical challenge for clinical ultra-high field (UHF) scanners (i.e. 7T), where multi-channel RF transmitter coils are used to enable RF shimming, and for preclinical UHF magnets with narrower bore sizes which constrain the transmit RF coil size and hence homogeneity. Although a combination of a parallel-transmit RF coil, RF shimming and multi-channel receive coil at 7T can produce good structural image quality in macaque monkeys (Gilbert et al., 2016), pial surface estimation errors can nonetheless occur in more inferior parts of the brain that are more distant from the surface coil transmit channels and consequently receive weaker $B_1+$ (Supp. Fig 2). To compensate for $B_1+$ bias, we recommend collecting a measure of $B_1+$ (without interpolation) even when MP2RAGE is used because excitation flip-angle may deviate from its nominal value and adiabatic condition may be violated due to limitations and inhomogeneities in $B_1+$ power (Haast et al., 2018). $B_1+$ variations due to dielectric effect in NHPs are modest at 7T relative to humans, but still benefit from RF shimming (Gilbert et al., 2016). The dielectric effect is even less for the marmoset brain at 9.4T, where $B_1+$ homogeneity is often dominated by the small RF birdcage coil or surface coil transmitters. Additionally, the center of the brain (i.e. thalamus) should be positioned close to the isocenter of the RF coil to minimize the $B_1+$ bias across the brain. For animals studied in a sphinx position with a body RF transmit coil, this can be challenging, but for studies where a custom, concentric RF coil is used (typically 7T and above) the head can be centered accordingly in the RF transmit field, where optimal uniformity is achieved, based on RF shimming solutions. It should, however, be noted that no amount of $B_1+$ correction can compensate for the loss of contrast when the RF pulses are not performing within their design parameters.

The receive RF field ($B_1-$) is intrinsically inhomogeneous when using multi-channel receive coils, as the sensitivity profiles decrease with distance from the individual receiver elements. This results in MR image signal intensity inhomogeneities which can be reduced using vendor provided prescan normalization (e.g. prescan normalize, Siemens). We highly recommend the use of receive biasfield corrections on the scanner for all images, as the receiver field is fixed to the head coil while the head moves around within it. Post-acquisition image-based corrections cannot readily replicate this effect and algorithmic corrections may identify real tissue inhomogeneities as "bias."

Prescan normalized images (Fig. 1a) may require additional intensity correction due to uncorrected $B_1-$ and shared $B_1+$ bias (Fig. 1b) to achieve robust tissue segmentation and generation of accurate cortical surfaces (Fig. 1c,d), for example if vendor algorithms are not optimized for smaller brains and head coils. Also, real tissue contrast (e.g. caused by differential grey and white matter myelination), may also impair segmentation algorithms that rely on within-tissue homogeneity. The signal intensity bias can be reduced during processing (e.g. using sqrt(T1w*T2w) approach in the HCP-NHP pipeline). Unfortunately, image types required for biasfield correction are sometimes not acquired in studies of NHPs. For example, in the PRIME-DE database (Milham et al., 2018), of the 16 imaging centers that provided structural MRI data, 10 centers provided both T1w and T2w images required for biasfield correction. Combining biasfield corrected and uncorrected structural data may have an adverse effect on the statistical power. Alternatively, we recommend acquiring structural images using a single-loop receive-only coil or bird-cage coil to substantially reduce the $B_1$- biasfield (Li et al., 2019; Zhu and Vanduffel, 2019) (Supp. Fig. 3), albeit this requires more averaging and a separate imaging session.

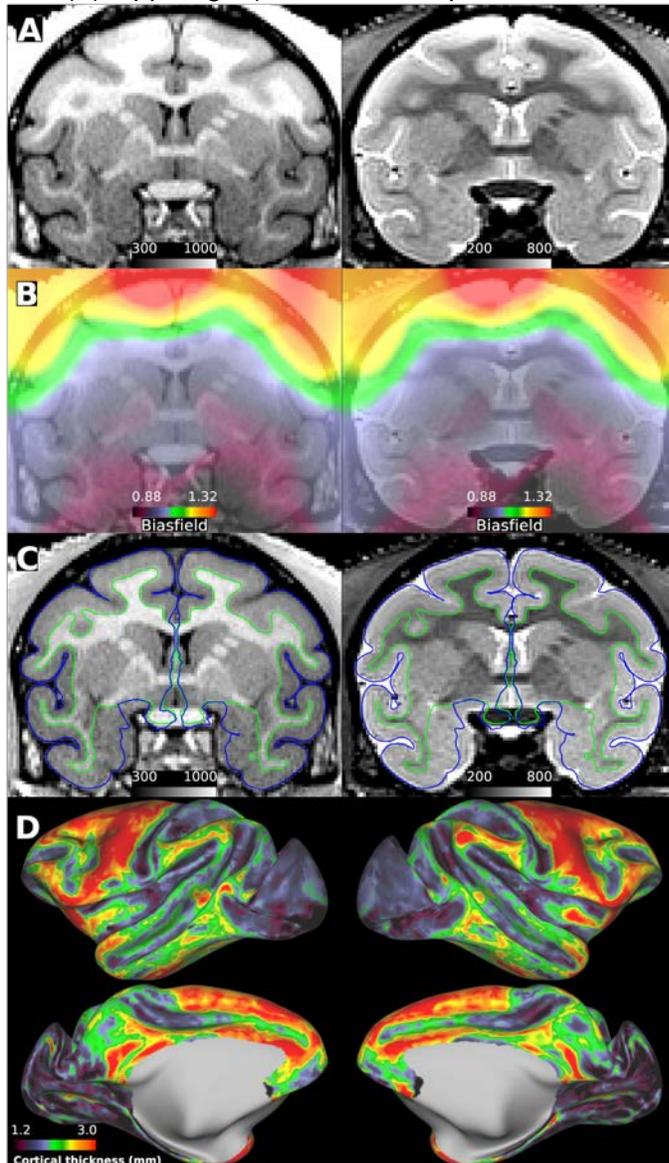

**Figure 1. Structural image quality standards and $B_1$- biasfield correction. (A)** Prescan normalized T1w MPRAGE (left) and T2w SPACE (right) images acquired with 0.5 mm isotropic resolution. Note the signal intensity bias near the superior surface of the brain despite prescan normalization and the decreased tissue contrast for myelin in the temporal lobes relative to the superior frontal lobes. **(B)** Intensity biasfield, due to uncorrected $B_1$- and shared $B_1$+, estimated using a square root of product of T1w and T2w images. **(C)** Biasfield corrected T1w and T2w images. The pial and white matter surfaces are indicated by blue and green contours, respectively. **(D)** Cortical thickness displayed over inflated cortical midthickness surface. Macaque data was acquired using the Human Connectome Project (HCP)–style data acquisition, preprocessed using non-human primate version of the HCP pipelines and visualized using HCP's Connectome Workbench (Autio et al., 2020).

To explore the variability of cortical thickness measures in macaque monkeys, we estimated cortical thickness maps using the HCP-NHP pipeline (Autio et al., 2020; Donahue et al., 2018; Glasser et al., 2013). We limited the data analysis to PRIME-DE centers that provided a high-resolution structural image and a $B_0$ fieldmap for distortion correction (see later reproducibility of resting-state FC). Across subjects (N=23), M132 atlas parcellated (Markov et al., 2014) cortical thickness varied between 1.4 mm and 3.4 mm with a mean 2.2 ± 0.5 mm (std) (Fig. 2a) and correlation coefficient (Pearson's) across subjects varied between 0.70 and 0.92 with an average 0.81 ± 0.02 (std) (Fig. 2b). Intra-site, parcellated cortical thickness correlations were RIKEN R=0.88 ± 0.05 (N=5), UC-Davis 0.87 ± 0.03 (N=5), MtS 0.86 ± 0.03 (N=5) and IoN 0.78 ± 0.06 (N=5) (Fig. 2d). Across the NHP imaging sites, correlation was also relatively strong (R=0.80 ± 0.06). IoN exhibited relatively lower correlation with respect to other imaging centers (R-values between 0.71 and 0.78), probably due to a lack of T2w image (other sites provided both T1w and T2w images). Moreover, cortical thickness distribution acquired using a multichannel surface transmit coil at 7T (Supp. Fig. 2a-d) was quite distinct from those acquired using volume transmit coils (R-values range between 0.33 and 0.60, N=2), suggesting a need for $B_1+$ field correction to improve detection of pial surface (Haast et al., 2020, 2018).

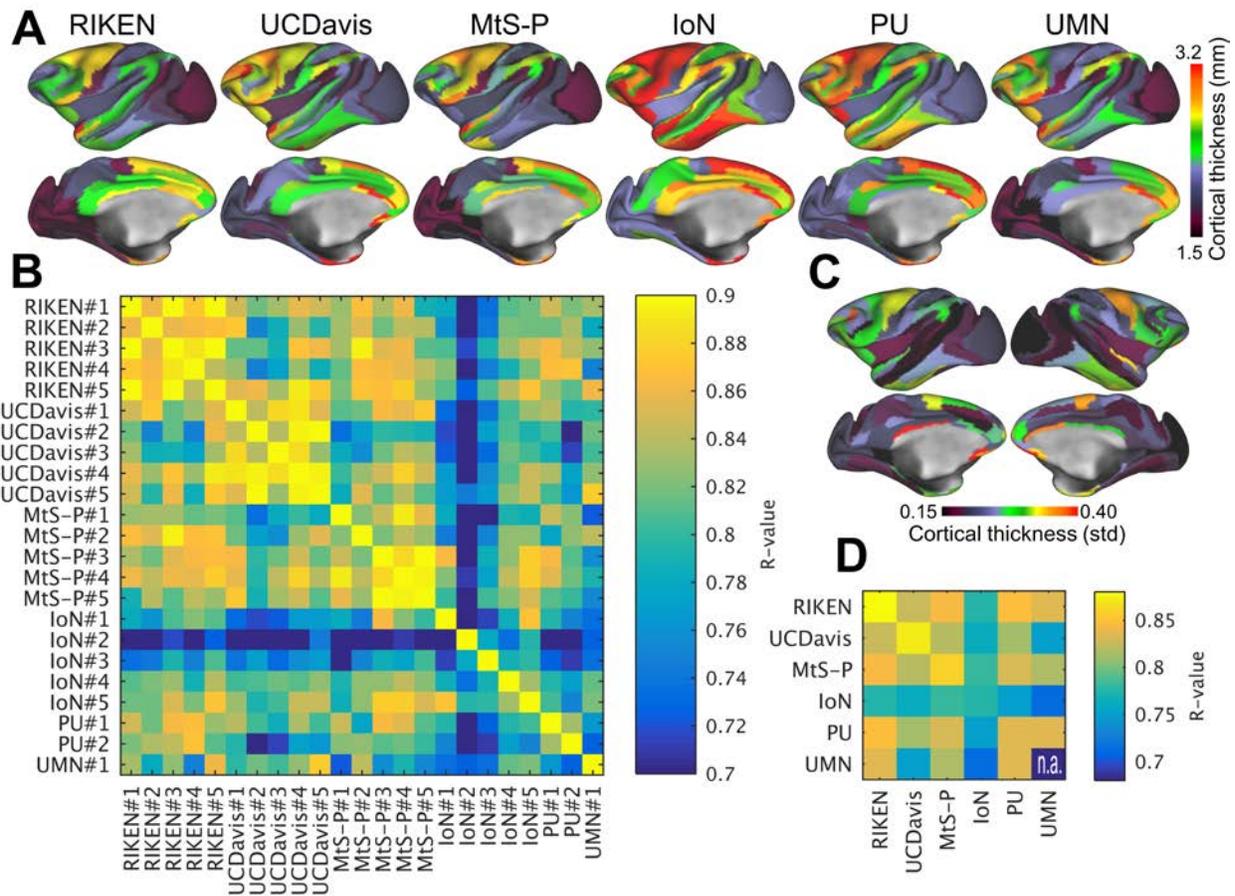

**Figure 2. Comparison of cortical thickness across exemplar subjects in the PRIME-DE. (A)** Top row shows exemplar parcellated curvature corrected cortical thickness maps from six PRIME-DE sites. **(B)** Comparison between cortical thickness maps across sites and subjects (N=23). **(C)** Variability of cortical thickness (N=23). **(D)** Average correlation across and within imaging centers. Cortical thickness maps were automatically generated using HCP-NHP pipelines (Autio et al., 2020; Donahue et al., 2016), parcellated using M132 atlas containing 91 parcels per hemisphere (Markov et al., 2014) and then Pearson's correlation coefficient between parcellated cortical thickness maps was calculated. Image resolution was 0.5 mm isotropic in all centers expect in UC-Davis resolution was 0.3 mm isotropic. Abbreviations RIKEN Institute of Physical and Chemical Research, Japan, UC-Davis University of California, Davis; MtS-p Mount Sinai-Philips; IoN Institute of Neuroscience; PU Princeton University; UMN University of Minnesota.

To corroborate the robustness of measurement of parcellated cortical thickness, we next analyzed test-retest data for five representative subjects (with subjects twice scanned within two months at RIKEN) and found an average R=0.97 ± 0.01 (N=5, for an exemplar test-retest pial and white matter surfaces and dense cortical thickness maps, see Supp. Fig. 4a-d and 5a-d). Reliability was comparable with YA-HCP (Supp. Fig. 6a-d). Taken together, the high reproducibility within subjects and variation of cortical thickness across subjects (at RIKEN dataset) suggests there is meaningful biological variation in the macaque population.

The extent of variation in regional cortical thickness due to measurement bias (i.e. environment, subject positioning, $B_1+$ and $B_1-$ biasfield, sequence protocol and segmentation) (Han et al., 2006) or biological differences (i.e. gender, body weight, head size, age and genetic) in the sampled multi-center macaque population remains unknown (Fig. 2c). To determine the primary sources of variation, harmonized data analysis is needed to explicitly remove center- and scanner-related effects from the multi-center data while preserving biologically-relevant covariates in the data (Fortin et al., 2018). A future challenge for the NHP community is to collect a sufficiently large range of biological measures across institutes for a large population of NHPs, through organized collaborative initiatives such as PRIME-DE, to identify biologically important factors with respect to the imaging data (i.e. brain volume, cortical surface area, regional cortical thickness and myelin) (Panizzon et al., 2009; Rakic, 1995; Schmitt et al., 2020).

2.2. $B_0$ field-map — preserving spatial fidelity

A $B_0$ field-map enables distortion correction of functional and diffusion (echo-planar) images so that they represent the physical space of the imaged animal. Unfortunately, currently $B_0$ field-maps are sometimes neglected in studies of NHPs. For example, in the open PRIME-DE database (Milham et al., 2018), of the 19 imaging centers that provided resting-state fMRI (rfMRI) data, only seven provided $B_0$ field-maps and only five provided EPI echo-spacing (time between neighboring *k*-space lines). Combining corrections for $B_0$ distortion and its interaction of motion may further increase spatial fidelity and statistical power (Andersson et al., 2001, 2003; Cusack et al., 2003).

Figure 3 demonstrates the geometric distortion of spin-echo (SE) EPI acquired with left–right (L-R) (Fig. 3a) and right-left (R-L) (Fig. 3b) phase encoding directions at 3T. Without distortion correction, the SE-EPI images remain poorly registered to (native) pial surfaces, but this was substantially improved by distortion correction using these opposing phase-encoding SE-EPIs (Fig. 3c,d). A shiftmap (Fig. 3e), which is estimated using a $B_0$ field-map and echo-spacing, demonstrates that the physical displacement of the imaging voxels in the cerebral cortex shifted up to 4 mm, with an absolute median of 1.1 mm. Because the macaque mean and minimum cortical thickness are approximately 2.1 mm and 1.0 mm, respectively, these spatial distortions require unwarping to precisely assign the fMRI voxels into grey matter and appropriate banks of sulci (Fig. 3a-d, arrow).

Field-maps can be estimated either from traditional multi-echo gradient-echo (GRE) images (using phase differences calculated from two, or more, TEs) or SE-EPI acquired in reversed phase encoding directions (either L-R & R-L or A-P & P-A traversal of *k*-space) (Andersson et al., 2003). However, SE-EPI $B_0$ field-map correction has additional benefits over GRE $B_0$ field-map corrections and non-linear registration-based methods (Andersson et al., 2003; Cusack et al., 2003; Graham et al., 2017; Holland et al., 2010) because it provides better SNR as a consequence of being less susceptible to signal-drop-out issues. In addition, SE-EPI can be acquired in a shorter time (i.e. one minute) than GRE field-map and is thus less susceptible to motion artefacts (important in awake NHP imaging). Importantly, using phase reversed gradient echo EPI is not recommended, as gradient echo images additionally have signal loss due to T2* dephasing that is mismatched between phase encoding directions and will be confused for geometric distortion by the internal registration algorithm used to estimate the $B_0$ field.

For the smaller NHP brain, FSL's TopUp $B_0$ field-map correction needs to be reconfigured from the default set-up, which is designed for the size of the human brain. We recommend adjusting the sampling resolution according to the isometric ratio between the NHP and human brains, as the scaling has a profound impact on the accuracy of distortion correction (Fig. 3f). After the TopUp configuration, the forward and reverse phase-encoded echo-planar images become more similar to each other, as measured by root-mean-square (RMS) deviation (mean RMS before and after configuration, 4596 and 1908, respectively; N=30), implying more accurate distortion correction.

Shimming can also reduce geometric distortion, and signal drop-outs, in EPI by improving $B_0$ homogeneity. However, automatized $B_0$ shim algorithms in the majority of clinical scanners are configured to the size of the adult human brain, which is approximately 10-fold larger in volume than the macaque, and 100-fold larger than the marmoset brain, respectively. Consequently, these algorithms are not necessarily effective in reducing susceptibility-induced $B_0$ inhomogeneities in small NHP brains (Fig. 3). An alternative approach to improve $B_0$ homogeneity is to manually fine-tune the shim coils or to use FastestMap (Gruetter and Tkáč, 2000), which enables adjustment of the length of line-scans to the spatial dimensions of the NHP brain.

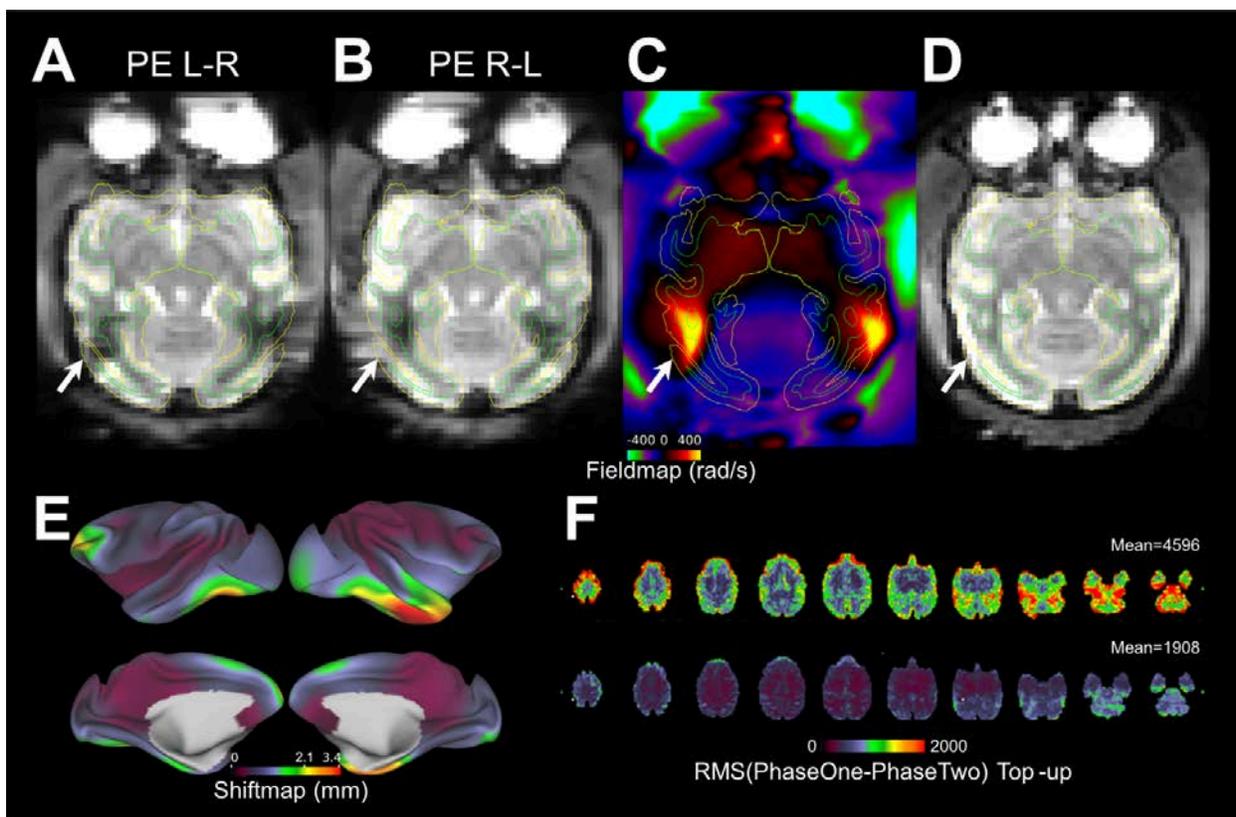

**Figure 3. Distortion correction is an important quality assurance standard to ensure spatial fidelity of functional and diffusion echo-planar images.** Functional single-band echo-planar images acquired with phase encoding (PE) directions **(A)** from left to right (L-R) and **(B)** from right to left (R-L). Note distortion, in particular near the temporal and occipital lobes (white arrows). **(C)** $B_0$ field-map, created using spin-echo (SE) echo-planar images. **(D)** Distortion corrected SE echo-planar reference image. Pial and white matter surfaces are indicated by the yellow and green contours, respectively. **(E)** Absolute shiftmap demonstrates the physical voxel dislocation (mm) due to magnetic field inhomogeneities. Shiftmap was calculated using FSL's utility FUGUE. **(F)** Root-mean-square (RMS) deviation of signal intensity between R-L and L-R PE echo-planar images before (top panel, mean 4596) and after (bottom panel, mean 1908) TopUp configuration for the size of the NHP brain (N=30).

Another primary means to reduce the geometric distortions in EPI is to reduce the echo-spacing. Practically, this can be achieved by utilizing strong gradients with fast slew rates or by choosing the phase encode direction in the orientation that allows shortest total readout time, which is determined by the required field-of-view (FOV) and the allowable echo spacing for a given direction. For example, using a L-R FOV would allow for a smaller phase FOV in comparison to the AP direction. In addition to shorter readout times this also enables a consequential reduction in TR. One caveat to this is that when using a commercial human system, peripheral nerve stimulation limits dictate the allowable echo spacings in any given gradient direction and tend to be more limiting in the L-R direction. Higher performing head only gradient inserts on human scanners can circumvent these limitations, significantly improving EPI image quality and efficiency.

Higher spatial resolutions in EPI are, however, particularly more challenging because of the longer required echo trains which result in increased signal-dropouts and resolution loss, in addition to the increased levels of geometric distortion. This is especially problematic at high $B_0$ and in regions with strong differences in magnetic susceptibility (or short T2*s), such as air-tissue interfaces. Such signal-dropouts and blurring can be effectively reduced using in-plane accelerations (i.e. Generalized autocalibrating partially parallel acquisitions GRAPPA). GRAPPA also enables to reduce echo-time (TE) further reducing artifacts, which is ideal for CBV measurements but less so for BOLD imaging as it reduces sensitivity. Yet another approach to shorten the effective TE is to use a segmented EPI acquisition with variable flip angles to normalize the intensity of each k-space segment, though this lengthens the TR. Therefore, segmented EPI yields excellent results in anesthetized but not in awake animals.

## 2.3. Functional MRI

### 2.3.1. Parallel imaging

The methodological development and application of parallel imaging have been on-going for the past two decades and have markedly improved human functional MR image quality (Griswold et al., 2002; Moeller et al., 2010; Pruessmann et al., 1999; Setsompop et al., 2012; Uğurbil et al., 2013; Wiggins et al., 2006). However, the translation of parallel imaging from humans to NHP has been nontrivial. The main barrier to translating cutting-edge pulse sequence protocols for NHPs has been the limited availability of dedicated multi-channel receive coils for NHP and the paucity of independent front-end channels in preclinical scanners. Notwithstanding the technical challenges, a growing number of NHP fMRI studies (Ekstrom et al., 2008) are demonstrating compelling benefits of parallel imaging yielding remarkable SNR gains used to improve spatial and temporal resolutions and statistical power (Box 2).

Slice direction acceleration using the MB technique has substantially improved the efficiency of functional 2D imaging in humans (Moeller et al., 2010), in particular towards higher isotropic resolutions with progressively thinner slice profiles. The MB RF transmission concurrently excites multiple slices that are unaliased using distinct multi-channel receive coil channel sensitivity profiles. Concurrent slice excitation enables a reduction in repetition time by the MB factor, however, the practical range of MB factors that can be achieved without inducing substantial cross-slice artefacts is constrained by the number of RF receive channels, their geometric arrangement and their corresponding sensitivity profiles (Cauley et al., 2014; Risk et al., 2018; Xu et al., 2013).

Recently, it has been demonstrated that a combination of macaque 24-channel coil and MB 5 provided the most efficient tSNR per unit time and reduced the TR to 0.7 sec from 3.8 sec (single band acquisition) with whole-brain coverage (Autio et al., 2020). The 5-fold increase in temporal data points enabled also more advanced multivariate analysis of fMRI timeseries in anesthetized macaques by providing an average 24 ± 11 (N=30) neural ICA components that would not have been possible using single-band BOLD fMRI at 3T. These findings are in line with more comprehensive observations made in humans using a 32-channel coil at 3T: MB 8, 0.7 s TR (compared to 5.7 s for a single band) provided most efficient tSNR per unit time (Smith et al.,

2013). The available data suggest the following crude guideline: each MB factor requires approximately four or five independent RF receive channels (HCP: 32-channel / MB 8 ≈ 4; macaque 24-channel / MB 5 ≈ 5). This of course depends on the actual levels of residual aliasing (and g-factor for GRAPPA), which should be independently measured for each coil design at a given field. Further, this also presumes that in-plane accelerations are not used (common only for 3T acquisitions), which would further limit the maximum MB factor as it also relies on information from coil sensitivity profiles. As such, the maximum acceleration would be limited by the combination of in-plane and through-plane accelerations (e.g. MB8 or MB5 x GRAPPA2) (Vu et al., 2017). This leads to a pressing need for high density coil systems for investigators interested in MB imaging in NHPs. High temporal resolution (i.e. ~1s or less TR) is critical for advanced data denoising techniques such as spatial and temporal ICA (Glasser et al., 2018) and improved statistics (Feinberg et al., 2010); however, we acknowledge that this may require new hardware for many sites and multi-band pulse sequences.

Acceleration capacity can be further improved by capitalizing on implanted phased-array coils that yield highly independent channel sensitivity profiles (see later, Increase sensitivity from implanted phased-array coils). Such coils (with 8 elements) enable further improvements in accelerated fMRI at 3T (8-channel / MB2×GRAPPA3≈1) (Janssens et al., 2012; Li et al., 2019; Zhu and Vanduffel, 2019).

---

**Box 2 │ Cost performance in NHP functional imaging**

**Acquisition hardware.** Contrast-to-noise ratio (CNR) is critical in fMRI. The primary means to improve CNR is to utilize scanners with ultra-high $B_0$ strength. Unfortunately, the narrow inner bore diameter (<30 cm) in the majority of preclinical scanners limits the possibilities of scanning large NHPs. However, the small size of the common marmoset makes them well suited for ultra-high $B_0$ scanners equipped with very strong gradients (i.e. 400 mT/m maximum gradient strength or up to 1,500 mT/m using gradient inserts) enabling high-resolution MRI (i.e. functional 0.5 mm isotropic) in conjunction with substantial increase in blood oxygen level dependent (BOLD) effect (Liu et al., 2020; Schaeffer et al., 2019; Silva, 2017). Another primary means to improve CNR is to utilize NHP dedicated multi-channel receive coils (Autio et al., 2020; Ekstrom et al., 2008; Gao et al., 2020; Gilbert et al., 2016; Janssens et al., 2013; Schaeffer et al., 2019; Yacoub et al., 2020) or phased-array coils embedded to the headcap of an animal (Janssens et al., 2012).

**Cost performance.** Viewed from a cost-benefit perspective, SNR supra-linearly increases with $B_0$ strength (SNR ~ $B_0^{1.65}$ (Pohmann et al., 2016)) for a cost of ~1 Million USD / Tesla (i.e. from 3T to 9.4T equates to ~7-fold increase in SNR for a cost of 9 Million USD). On the other hand, a 24-channel RF receive coil provides ~5-fold tSNR increase in macaque monkeys and a 16-channel coil provides ~20-fold tSNR increase in common marmosets at 3T for a cost of ~0.1 Million USD (tSNR increase in cerebral cortex was estimated relative to the human connectome project at 3T and adjusted according to the voxel size) (Autio et al., 2020). Moreover, implanted phased-array coils yield further SNR increases (~9 × relative to 8-channel external coil) in macaques at even lower cost (~0.01-0.05 Million USD) (Janssens et al., 2012), albeit each subject needs to be surgically equipped with its own set of coils which is not feasible in many laboratories. Multichannel receive coils also enable adapting accelerated image acquisition protocols (MB and GRAPPA) yielding further improvements in SNR per unit time. Thus, this putative cost performance analysis favors prioritizing multi-channel RF receive coils, albeit the acquisition of hardware is complicated by the fact that the scanner purchases are often supported by institutions whereas receive coils are typically acquired using laboratory-specific resources. A complementary cost effective means to improve fMRI is to capitalize on contrast agents (i.e. MION) for a ~3-fold CNR increase in task fMRI at 3T for a cost of ~100 USD / scan session (Vanduffel et al., 2001).

**Prospects.** A combination of ultra-high $B_0$ (e.g. clinical 7T or 10.5T Siemens or preclinical Bruker 9.4T), gradient insert and a phased-array receive coil, and possible with additional use of contrast agents, is expected to provide the state-of-the-art experimental setups. Extrapolated, such combination predicts an over 10-fold increase in macaque SNR (($B_0$ ratio)$^{1.65}$ × multichannel coil gain=(7T / 3T)$^{1.65}$ × 5 ~ 20) and an over 100-fold increase in marmoset SNR ((9.4T / 3T)$^{1.65}$ × 20 > 100), while not taking into account the gains from increased BOLD effect, reduced partial volume effects and improved accelerated imaging. Implanted 8-channel phased-array coil in macaque monkeys predicts further increase in cortical SNR ($B_0$ ratio × gain vs external 8-channel coil=((7T / 3T)$^{1.65}$ × 9 ~ 30). This speculation suggests that prospects to improve upon our current understanding of functional organization in NHP, up to laminar and columnar level are, indeed, very promising.

---

While ultra-high $B_0$ provides further sensitivity gains for fMRI and improved parallel imaging performance it also necessitates in-plane accelerations. Gilbert and colleagues used a combination of 24-channel RF receive coil, MB2×GRAPPA2 at 7T (Gilbert et al., 2016) whereas Yacoub and colleagues used a combination 32-channel RF receiver coil, MB2×GRAPPA3 at 10.5T (Yacoub et al., 2020), which may yield improved sensitivity to neuronal activity. However, to maintain animal safety standards, care should be taken not to exceed specific absorption rate (SAR) safety regulations (level 1) at ultra-high $B_0$.

In awake behaving NHPs, we recommend paying special attention to the quality of the GRAPPA auto calibration scan, used for accelerated image reconstruction, as it may be compromised by physiological noise or motion during the acquisition of the reference and motion between the reference and the fMRI acquisitions. To compensate for the former issue, we recommend to acquire ~30 sec GRAPPA auto calibration data using GRE FLASH, rather than single-shot or segmented EPI, to average physiological noise and to improve temporal SNR (tSNR) (Polimeni et al., 2016; Vu et al., 2017). For the second issue, on-line supervision of the animal's behavior during the reference scan is recommended to ensure that no substantial motion biases the quality of the GRAPPA auto calibration data. NHPs which are well trained and acclimatized to the MRI environment are substantially less susceptible to such calibration artefacts that can have dramatic influences upon image reconstruction. Motion control is also important between the GRAPPA reference acquisition and the fMRI time series, otherwise reconstruction errors can occur. An alternative strategy is to acquire multiple GRAPPA references and then use the one that is closest in time to the particular fMRI run.

However, single-piece external multi-array coils may hamper the accessibility of electrophysiological, microstimulation, optogenetic or two-photon devices, which are invaluable tools to explore the underpinnings of functional organization in NHPs. The external RF designs can be specifically designed with specific openings that permit insertion of electrodes (Gilbert et al., 2018; Schaeffer et al., 2019), albeit with limited access to the different brain regions. An alternative powerful (but technically demanding) option to mitigate this problem is to implant receive coils directly above the skull yielding improved signal due to reduced tissue-coil distances (see later Increase sensitivity from implanted phased-array coils) (Janssens et al., 2012).

2.3.2. Spatial resolution

To reduce partial volume effects (between grey matter and whiter matter and CSF) and to distinguish between opposing banks of sulci, we recommend adopting the HCP strategy of adjusting the (isotropic) fMRI voxel resolution according to the thickness of cerebral cortex (median thickness 2.1mm in the macaque) (Glasser et al., 2016b, 2013; Yacoub et al., 2003). Thus in macaque monkeys, Autio and colleagues adjusted the fMRI resolution (1.25 mm) below to the lower 5th quantile of cortical thickness (1.4 mm) at 3T (Fig. 4a) (Autio et al., 2020) whereas Gilbert and colleagues adjusted the fMRI resolution to the thinnest parts of the cerebral cortex (1.0 mm) at 7T (Fig. 4b) (Gilbert et al., 2016). In humans, higher resolution (1.6 mm) acquired at 7T provides substantially improved functional CNR with less partial volume effects and reduced cross-sulcal artefacts (Vu et al., 2017; Yacoub et al., 2003) and similarly UHF studies of NHPs may yield spatially more fine-scaled localization of brain functions (Schaeffer et al., 2019; Yacoub et al., 2020). We acknowledge that hitting both the spatial resolution and temporal resolution targets will require multi-band sequences with multi-channel head coils that may not be currently available at all sites.

In NHPs, however, implanted RF coils enable much higher spatial resolution, even at 3T (Fig. 4c, d) (Janssens et al., 2012; Li et al., 2019; Zhu and Vanduffel, 2019). Such high resolution offers great potential to investigate the mesoscale columnar and laminar organization of the macaque cortex, even in cortical areas where accessibility is currently not possible with microscopic imaging tools.

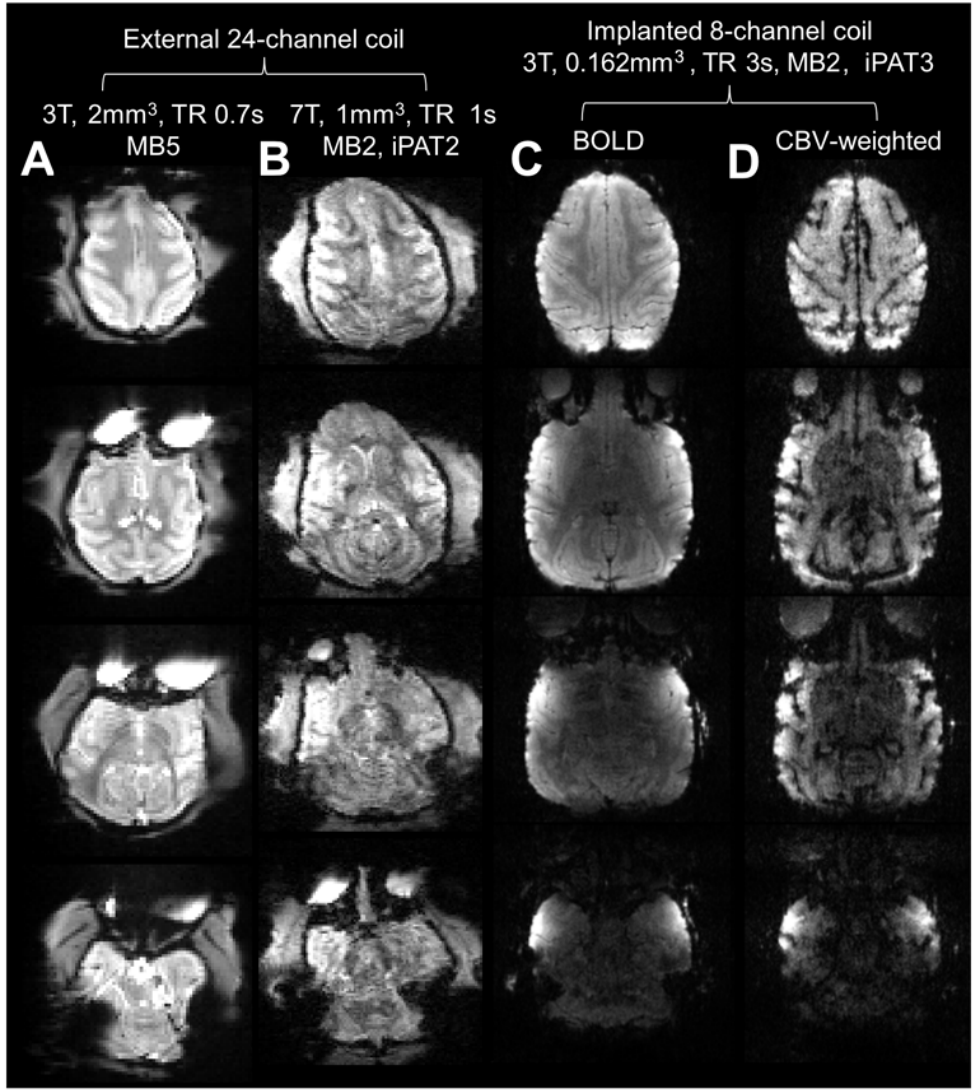

**Figure 4. Comparison of echo-planar image quality across different hardware configurations.** Single blood oxygen level dependent (BOLD) echo-planar images acquired on anesthetized macaque monkeys using a 24-channel coil **(A)** at 3T (Autio et al., 2020) and **(B)** at 7T (Gilbert et al., 2016). Echo-planar image quality may be further improved using implanted phased-array coils at 3T with **(C)** BOLD or **(D)** cerebral blood volume weighted (CBVw) fMRI (Janssens et al., 2012). Note that implanted RF coils enable an order of magnitude smaller voxel size in comparison to conventional multi-channel coil designs while maintaining a good signal-to-noise ratio at majority of the cortical surface. Although echo-planar image quality is an important requirement, we emphasize that it is only one factor towards acquiring high sensitivity to neuronal activity.

As for EPI *k*-space coverage, to maintain spatial fidelity of the images, we recommend acquiring as much as possible of *k*-space because partial-Fourier encoding, which omits outer *k*-space lines that represent high spatial frequencies, reduces the spatial fidelity of functional and diffusion MR images (i.e. it blurs). In practice, full *k*-space coverage is feasible using preclinical scanners equipped with very strong gradients (i.e. ~15-cm-diameter coil with a maximum gradient strength of 400 mT/m (Schaeffer et al., 2019)), however, using clinical scanners equipped with relatively weaker gradients (e.g. Siemens, Trio and Prisma with a maximum gradient strength of 40 mT/m and 80 mT/m, respectively) compromise, in the form of less *k*-space coverage (reducing resolution) or interleaved acquisitions (reducing TR) may be required to avoid substantial signal dropout during read-out, in exchange for more complete brain coverage. Finally, it should be noted that for most clinical scanners software limitations prohibit users from employing the full strength of the gradient coils per unit time (dB/dt) for safety reasons and for protecting the equipment. Using the equipment at full strength, however, can be crucial for achieving high spatial or temporal resolution imaging. Stringent research agreements with the vendors are typically required to disable such built-in safety measures.

### 2.3.3. Length of imaging session and temporal resolution

Length of the rfMRI data acquisition is another important consideration for obtaining high quality FC (Birn et al., 2013; Laumann et al., 2015; Xu et al., 2018). The relationship between FC (Z-transformed correlation coefficient) and scan duration is evident using a seed point and the rest of the cerebral cortex in a representative anesthetized (isoflurane 0.6% and dexmedetomidine 4.5 µg/kg/hr) macaque monkey (Fig. 5a) and (awake) human (Fig. 5b) (Smith et al., 2013). Clearly, Z-scores (both positive and negative anti-correlations) span a wider range for longer scan durations, demonstrating the statistical gain achieved through the accumulation of temporal time-points (Fig. 6a,b) while revealing neurobiologically meaningful FC patterns (Fig. 5). Interestingly, FC distributions become more skewed for longer scan durations (Fig. 6), in line with the skewed distribution of cortico-cortical connection weights (Markov et al., 2014).

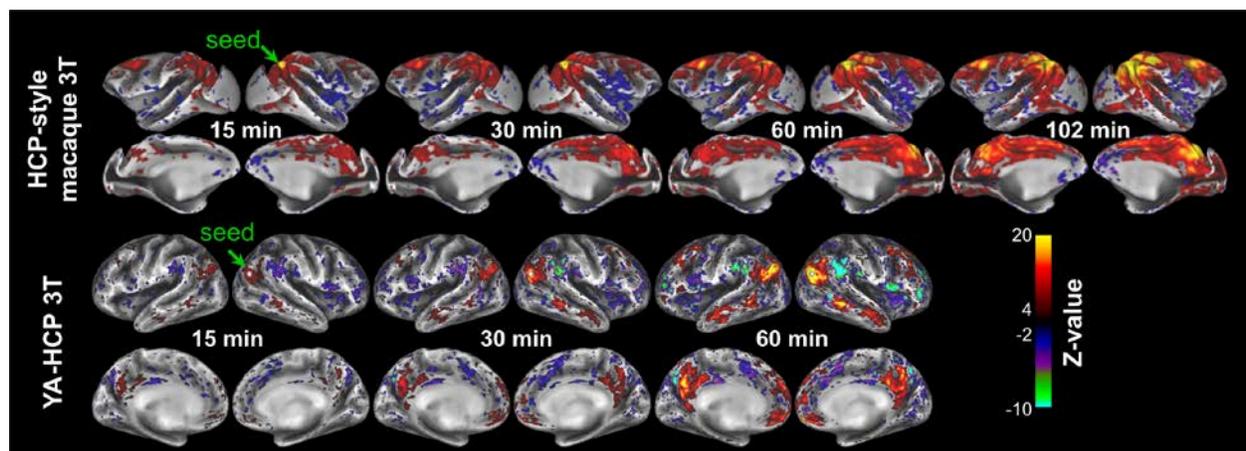

**Figure 5. Longer resting-state fMRI scan duration improves the quality of functional connectivity metrics.** Functional connectivity (Z-transformed Pearson's correlation coefficient) between a seed point (single grayordinate seed) in the default mode network area and the rest of the cortical mantle. In the macaque two sessions each 51 minutes are acquired whereas in the Young Adult Human Connectome Project (YA-HCP) four sessions were acquired with each 15 min length. Both macaque and human BOLD fMRI data were acquired with repetition time ≈0.7 sec (Autio et al., 2020; Smith et al., 2013) and data was preprocessed using HCP and non-human primate (NHP)-HCP pipelines (Autio et al., 2020; Donahue et al., 2018; Glasser et al., 2013), including FreeSurfer (Fischl, 2012) and ICA-FIX processing (Griffanti et al., 2017, 2014; Salimi-Khorshidi et al., 2014).

In the PRIME-DE database, rfMRI scan length varies between 8 min and 155 min per subject, while repetition times range between 0.7 sec and 2.6 sec (Milham et al., 2018). These, and other factors (i.e. receiver coil, imaging parameters, and anesthesia protocol), result in resting-state FC Z distributions that vary widely across sites (Fig. 6b, c).

The length of the imaging session and number of temporal data points also has implications for data-driven analyses of fMRI timeseries. For example, ICA separates multivariate fMRI timeseries into subcomponents that are non-Gaussian and statistically independent (Beckmann et al., 2005; Beckmann and Smith, 2004). However, to distinguish between unstructured noise (Gaussian distribution) and structured components (non-Gaussian distribution) the fMRI timeseries variance needs to be appropriately sampled. Indeed, the number of spatially independent components increases with respect to the length of the imaging session (with a linear coefficient of 2.4 components / min, Fig. 7b), as determined by FSL's Multivariate Exploratory Linear Optimized Decomposition into Independent Components (MELODIC) software. The majority of these sICAs, however, are time-varying imaging artefacts (i.e. motion, respiration and MR-artefacts) that need to be removed from the fMRI timeseries to obtain neurobiologically meaningful FC profiles (Fig. 7).

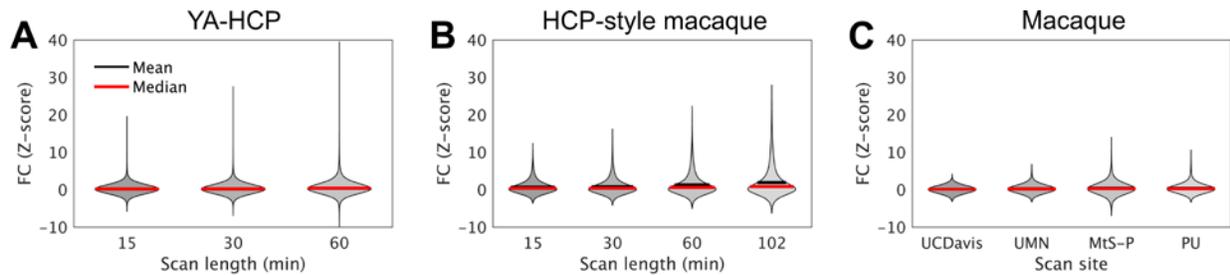

**Figure 6. Distribution of seed-based resting-state functional connectivity (FC; Z-transformed correlation coefficient) in cerebral cortex.** **(A)** The young-adult human connectome project (YA-HCP) (Smith et al., 2013) (Subject ID: 100307). **(B)** HCP-style macaque imaging (Autio et al., 2020). **(C)** Representative PRIME DE-sites (Milham et al., 2018). Data was distortion corrected, detrended, motion corrected and FIX-cleaned using HCP-NHP pipelines (Autio et al., 2020; Glasser et al., 2013). Violin plots contain approximately $60 \times 10^3$ nodes and $1.8 \times 10^9$ edges in YA-HCP whereas they contain approximately $18 \times 10^3$ nodes and $160 \times 10^6$ edges in macaque monkeys. Local FC (2% geodesic distance) is not shown. Scan length and number of volumes acquired were 13 min and 500 at University of California, Davis (UC-Davis), 27 min and 1,600 at University of Minnesota (UMN), 42 min and 8,192 at Mount Sinai School of Medicine (Philips) (MtS-P) and 60 min and 1,824 at Princeton University, respectively.

One popular means to remove the structured time-varying artefacts from fMRI timeseries is to use model-free FMRIB's ICA-based X-noiseifier (FIX) (Griffanti et al., 2017, 2014; Salimi-Khorshidi et al., 2014; Smith et al., 2013). Such structured artefacts account for 50% more variance than neural BOLD in anesthetized macaque monkeys at 3T (Autio et al., 2020). Typically, the largest noise components occur at ventilation frequency (Fig. 7a), reflecting subtle respiration-related head movements, and/or modulation of $B_0$ by respiration or body motion, that cause spurious long-distance correlations (Fig. 7c,e) (Fair et al., 2020; Power et al., 2012; Teichert et al., 2010). After removal of such nuisance artefacts using ICA-FIX, artefacts are profoundly reduced (Fig. 7e,f) and seed-based FC exhibits mainly short distance connectivity (Fig. 7d). Techniques like sICA+FIX benefit greatly from high temporal resolution (~1s or less TR) and long acquisition runs or combining across multiple runs (Glasser et al., 2018). Pre-scan normalization is also helpful in reducing motion artefacts that come from the head moving around within a static receive field.

Taken together, the length of a functional imaging session and temporal resolution have implications for removing unwanted nuisance signals from the fMRI timeseries and increasing statistical significance. We advocate fMRI imaging session durations that are at least one hour for anesthetized animals, whereas in awake imaging the scan duration should be determined according to the well-being and performance of an animal in each experimental setup, noting that time series can be concatenated from multiple sessions. In anesthetized animals, blood pressure and heart rate should be continuously monitored because prolonged anesthesia sessions tend to reduce blood pressure which in turn is compensated by an increase in heart rate. Future studies establishing scan lengths that maximize within animal replicability are needed (Laumann et al., 2015; Xu et al., 2018).

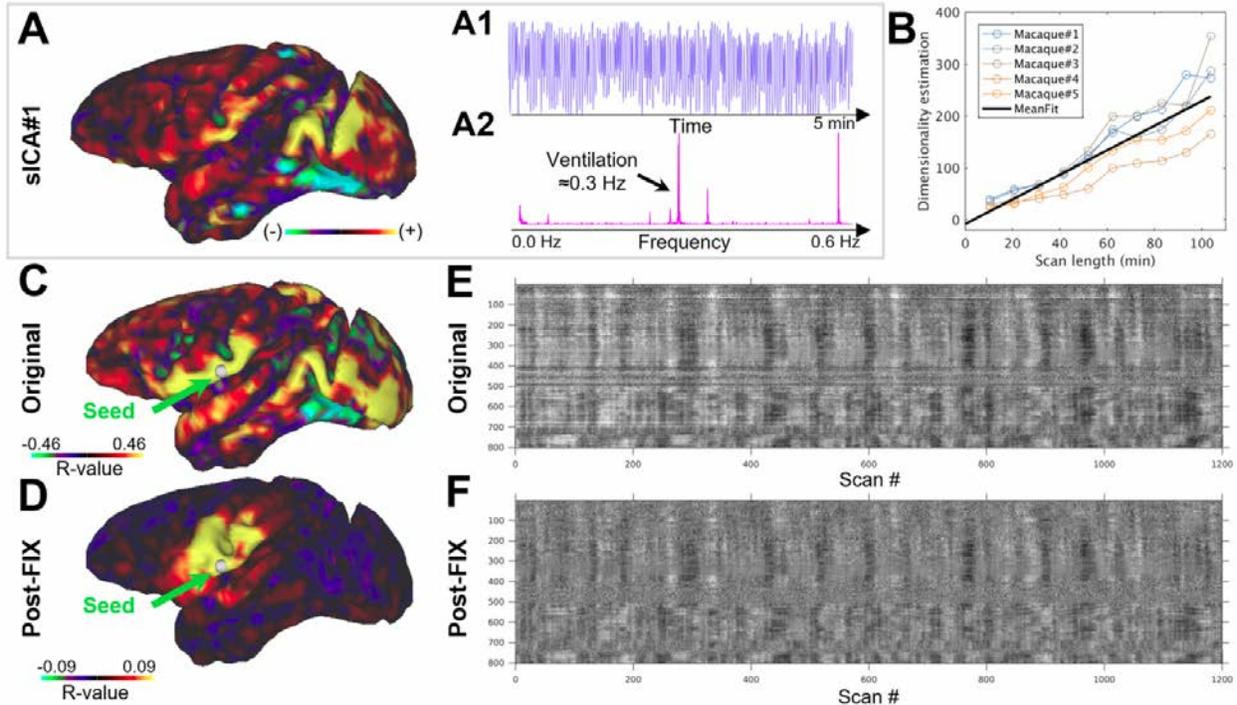

**Figure 7. Quality assurance analysis of functional timeseries and functional connectivity. (A)** Typical MRI artefact in an anesthetized macaque monkey identified using spatially independent component analysis (sICA). Note that this artefact exhibits **(A1)** temporal oscillations **(A2)** at ventilation frequency. **(B)** The number of identified sICAs (including both noise and neural networks) increases with respect to the scan duration. **(C, D)** Comparison of functional connectivity between a seed point in area 2 (green arrow) and the rest of the cortical mantle **(C)** before and **(D)** after FIX-ICA clean-up. Note that before fMRI preprocessing there are large spatially specific signal fluctuations and FC do not appear neurobiologically meaningful whereas after the clean-up these fluctuations are reduced and strong functional connectivity is dominated by neurobiologically sensible connections. Grayplot of **(E)** uncleaned (but distortion corrected) and **(F)** FIX-ICA cleaned (including motion correction, detrending and FIX clean-up) (Power, 2016; Power et al., 2014). The grayplots are scaled according to % parcel mean signal (±2%) balanced according to parcel size (Markov et al., 2014) and for visualization purposes are ordered by hierarchical clustering (Ward's method) (Glasser et al., 2018). Note the reduction of spatially specific fluctuations (horizontal bands) after FIX-cleanup.

### 2.3.4. Reproducibility of resting-state functional connectivity

The open PRIME-DE data repository offers a unique opportunity to examine the reproducibility of NHP neuroimaging to advance towards a scientific consensus on the functional organization of NHPs (Milham et al., 2018; Zuo et al., 2014). Here, we explore reproducibility of anesthetized macaque resting-state FC matrices at three levels: within a session (split-half scan similarity), within imaging centers, and across centers. The data analysis was limited to PRIME-DE centers that provided a high-resolution structural image and a $B_0$ fieldmap for geometric correction. Split-scan reproducibility of M132 atlas parcellated (Markov et al., 2014) FC matrices calculated using FIX-denoised fMRI data was high in all centers: RIKEN rho=0.94 ± 0.04 (N=5), UC-Davis 0.82 ± 0.11 (N=5), MtS-P 0.97 ± 0.02 (N=5), IoN 0.91 ± 0.03 (N=5), PU 0.87 ± 0.02 (N=2), and UMN 0.89 (N=1) (throughout the text FC is reported in units of Spearman rank correlation rho). However, intra-site reproducibility of parcellated FC matrices across subjects was variable: RIKEN rho=0.61 ± 0.05, UC-Davis 0.39 ± 0.06, MtS 0.28 ± 0.10, IoN 0.25 ± 0.15 and PU 0.50 (Fig. 8a, b). Across the NHP imaging sites, reproducibility of parcellated FC was strikingly low (0.23 ± 0.13; rho ± std). Best, yet weak, reproducibility was found between RIKEN and PU (rho-values ranging between 0.33 and 0.41; mean 0.36 ± 0.03).

Test-retest reliability of Z-scored parcellated FC matrices was promising in RIKEN (Fig. 8c). The reproducibility was relatively weaker in regions with weaker tSNR and cortical areas that are smaller in size (Fig. 8d), as expected.

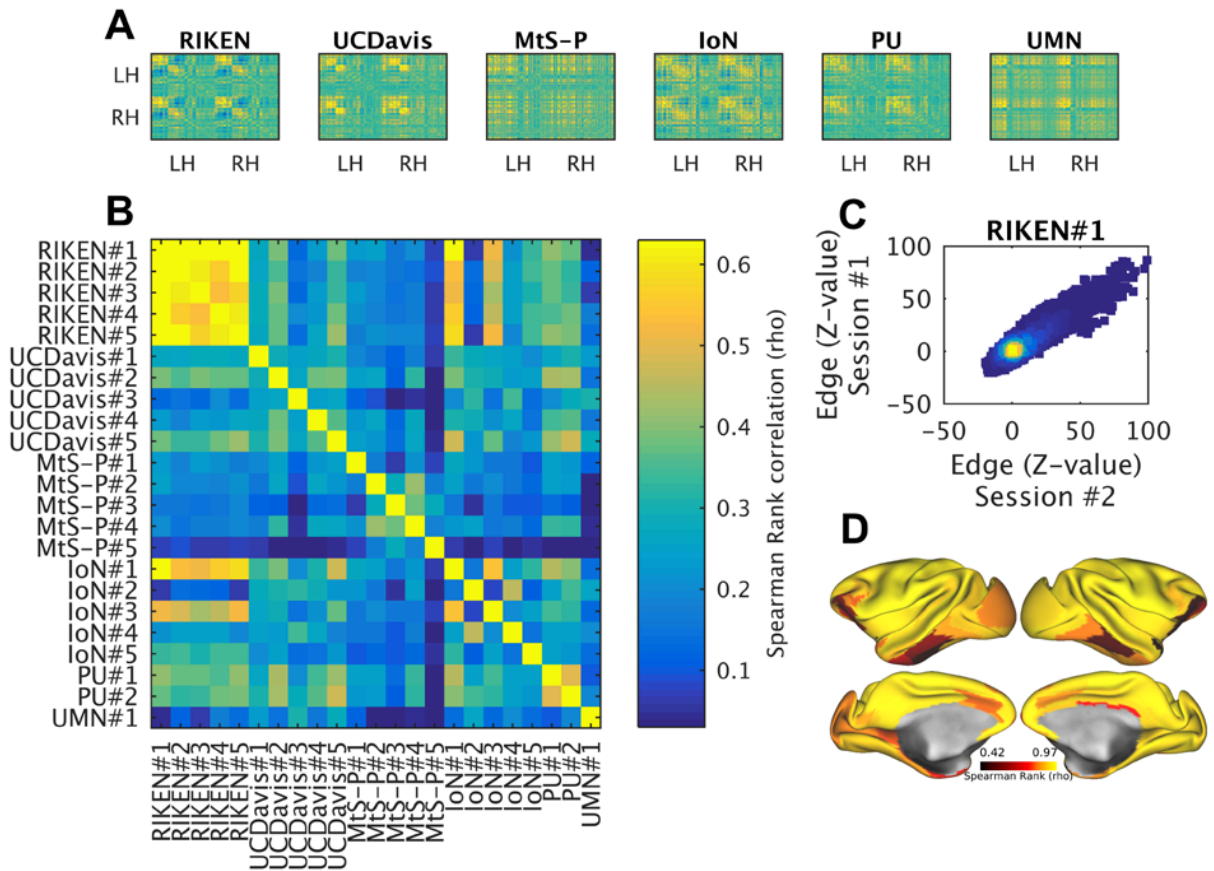

**Figure 8. Reproducibility of resting-state functional connectivity (FC) within and across PRIME-DE macaque imaging sites.** **(A)** Exemplar FC correlation matrices from six PRIME-DE sites. Representative matrices closest to global average are shown and ordered according to hierarchical clustering (Ward's method). **(B)** Comparison between correlation matrices across sites (six) and subjects (total N=23). fMRI timeseries were preprocessed using HCP-NHP pipelines, parcellated using M132 atlas containing 91 parcels per hemisphere (Markov et al., 2014), and then Spearman's Rank correlation coefficient (rho) between parcellated timeseries was calculated. Comparison of FC was limited to PRIME-DE sites that fulfilled minimum acquisition criteria (high-resolution anatomical image and a $B_0$ field-map). **(C)** Test-retest (heat) scatter plot of Z-scored FC matrixes (N=1, n=2, RIKEN data). **(D)** Reproducibility was high (>0.8; rho) in majority of the cortex (>78%), however, areas distant to RF receive channel coils and weaker SNR (i.e. hippocampal complex and ventral visual stream) exhibited lower reproducibility (RIKEN data was acquired using HCP-style protocols). Abbreviations: HCP the human connectome project; UC-Davis University of California, Davis; MtS Mount Sinai-Philips; IoN Institute of Neuroscience; PU Princeton University; UMN University of Minnesota, RH right hemisphere; LH left hemisphere.

### 2.3.5. Task specifications

A prerequisite for acquiring robust imaging data from NHP during task performance in the scanner relates to extensive training and behavior of the animal and the acclimatization to the scanning procedures (with recorded scanner noise and mock receive coils). For simple sensory stimulation experiments, such as a passive viewing, auditory, or tactile experiment, it is recommended to train the animals so that they can fixate for a high proportion of the scan session (e.g. > 90% of a typical scan duration), while reward delivery is also contingent upon fixed hand positions of the animal. The combined control of eye gaze and hand position dramatically reduces body motion and motion-related artifacts, thereby improving the quality of the EPIs. Physically restraining the body of the animals may have a counterproductive effect, as they (e.g. macaque) may resist and attempt to move more than without restraint. However, this varies across species, and some species (e.g. marmoset) respond well to body confinement (Silva, 2017).

For both block and event related designs, imaging protocols need to be adapted for BOLD or CBVw fMRI: the hemodynamic response function (HRF) in CBVw fMRI is more prolonged in comparison to BOLD (Leite et al., 2002; Vanduffel et al., 2001).

2.3.6. Contrast agents

Many NHP imaging sites use contrast enhanced imaging to amplify CNR in fMRI (Vanduffel et al., 2001). In particular, monocrystalline iron oxide nanoparticles (MION) have been used to amplify cerebral blood volume weighted (CBVw) variance in fMRI timeseries.

Early studies in rodents demonstrated improved sensitivity of CBVw over BOLD fMRI (Mandeville et al., 1998), and follow up studies in macaques have shown that using a dose of 8-10 mg/kg of MION increases CNR by a factor of ~5 at 1.5T and by ~3 at 3T (Leite et al., 2002; Vanduffel et al., 2001). Additionally, CBVw fMRI shows no large vessel artifacts as opposed to BOLD, because adjacent to large vessels the strong susceptibility gradients induced by MION attenuate the majority of the T2*-weighted MR signal. Importantly, the relative amplitude of CBVw response peaks in parenchymal brain tissue (Zhao et al., 2006), whereas the BOLD effect is biased towards large superficially located draining veins (Fig. 9). Early work with MION contrast agents suggested that due to the longer recovery time of the MION signal (Vanduffel et al., 2001) shorter stimulus periods might not benefit as much from the use of iron-based contrast agents. However, Leite et al (Leite and Mandeville, 2006) showed that even for event related designs, CBVw fMRI provided a significant increase in CNR over BOLD fMRI, though the increase was lower than for block designed paradigms. Moreover, recent studies have also shown that resting-state FC is more reliably measured using CBVw than BOLD fMRI (Xu et al., 2018). Taken together, for the majority of NHP fMRI studies we recommend the use of MION to increase CNR and statistical strength. This statistical gain is evident in an exemplar task activation study acquired using BOLD and CBVw fMRI (Fig. 9).

However, the use of MION also comes with a health burden to the animal that requires careful monitoring and countermeasures to maintain the animal's well-being. In particular, excess iron is known to accumulate in liver, spleen, lymph nodes, lungs and fatty tissue, which can, in the long run, lead to steatohepatitis, fibrosis, cirrhosis, hepatocellular carcinoma, and other diseases. Thus, regular monitoring of blood iron levels is necessary and should be part of protocols that use MION or other iron-based contrast agents for functional imaging.

Especially after repeat dosing, but even after a single high dose of MION, iron deposits can be found in the brain where they negatively impact imaging by causing signal dropout. Although mechanisms for iron uptake into neurons and astrocytes exist, studies in rodents suggest that these deposits primarily occur in the choroid plexus (Gorman et al., 2018), and there have been no reports of neurological side effects so far. Unfortunately, these deposits can persist for extended time periods, thus potentially rendering animals unusable for further imaging experiments. However, build-up of iron deposits can be prevented and existing deposits can be released to some extent by using iron chelators (Vanduffel et al., 2001). Apart from the commonly used Desferal, which can be conveniently administered by injection, Deferiprone is a particularly promising counter-measure. In contrast to Desferal, which mostly targets the liver and spleen and does not cross the blood brain barrier (Mounsey and Teismann, 2012), Deferiprone reduces brain iron accumulation without interfering with normal brain iron signaling (Boddaert et al., 2007; Mounsey and Teismann, 2012) and seems to be safe in monkeys even when administered daily for up to a year (Connelly et al., 2004). Both compounds can also be used safely in combination, but care should be taken not to over-chelate which can result in iron deficiency. Therefore, measuring ferritin, transferrin and iron levels at regular intervals is recommended to achieve a safe chelation protocol in individual monkeys.

## 2.4. Increased sensitivity from implanted phased-array coils

Implanted phased-array coils allow for substantial increases in the sensitivity of fMRI, even at relatively low $B_0$ strengths (Janssens et al., 2012; Li et al., 2019; Zhu and Vanduffel, 2019). At 3T, implanted array coils, with as few as eight elements, enabled functional imaging of the entire macaque brain using accelerated imaging under awake conditions at 0.6 mm isotropic resolution (with tSNR between 40-60 in the cerebral cortex) (Fig. 4c, d). This approach was used to reliably resolve fine-grained functional compartments such as the V2 stripes (Li et al., 2019). Such resolution is over 2-fold higher than the state-of-the-art isotropic resolution achieved in humans at 7T (0.8 mm isotropic), and relative to median cortical thickness is comparable across species (resolution / median cortical thickness: macaque 0.6/2=0.29 and human 0.8/2.7=0.30). Resolution can be further improved by increasing the number of channels and applying the same technology for NHP at 7T or higher $B_0$ strengths (Box 2).

The phased-array coils can be embedded in the headset of monkeys, typically used to fix their skull during awake experiments, which minimizes the distance relative to the brain and yields a fixed loading of the coils across scanning sessions. However, each subject requires a unique set of implanted coils. In particular, the loop array needs to be constructed to fit the unique shape of the skull of each individual subject to improve SNR, a subject-specific set of external matching circuits needs to be constructed according to the unique sizes of the loops, and invasive surgery is needed to embed the elements in the headpost of the animals, just above the skull. Another challenge of implanted arrays is that access ports to the brain (e.g. recording wells) are more restricted in their size and location and are preferentially planned before implanting the coil.

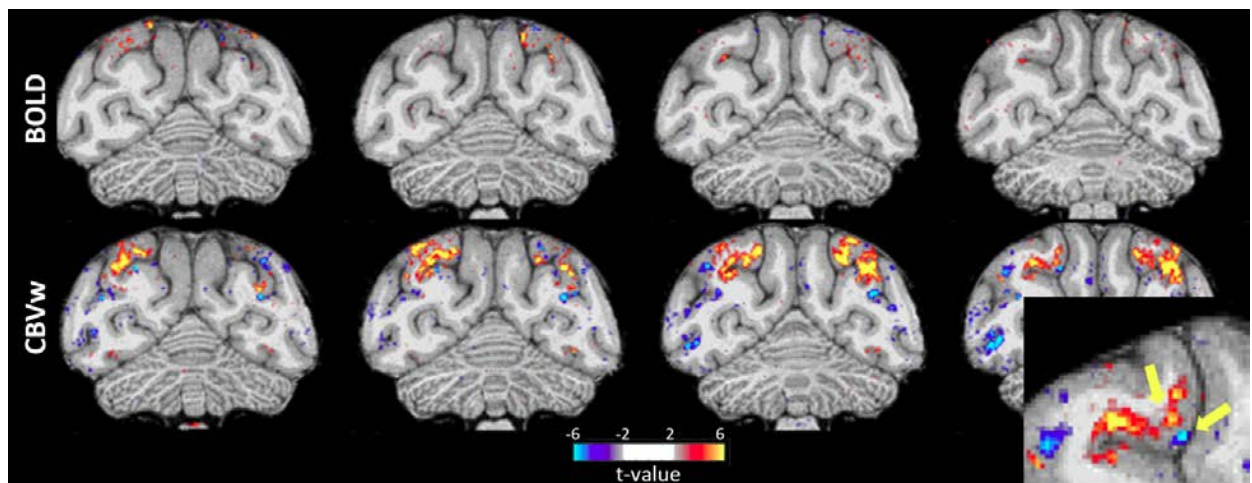

**Figure 9.** Comparison of blood oxygen level dependent (BOLD; top row) and cerebral blood volume weighted (CBVw, MION; bottom row) fMRI activation maps of viewing scenes versus objects obtained from the same subject with an implanted phased-array coil on consecutive scan days. The same number of runs from a single imaging session with equal fixation performance (> 90% within a 2×2° window) were used for the analysis. Note that CBVw exhibits much higher sensitivity than BOLD (t > 2). BOLD signals are typically highest at the pial surface (draining veins) whereas the CBVw activation maps reveal differential responses in upper versus lower layers (see yellow arrows in lower right enlarged panel). Data was acquired using standard gradient-echo EPI sequence at 3T Siemens Prisma scanner (0.162 mm$^3$ voxels, TR 3 s, MB 2, GRAPPA 3 and volumes 220).

## 2.5. Diffusion MRI

The minimal specifications for dMRI are less well established due to uncertainty in required spatial resolution to resolve underlying white matter fiber architecture (Donahue et al., 2016; Liu et al., 2020), or microstructural properties in grey matter (Autio et al., 2020; Fukutomi et al., 2018). Subsequently, the spatial resolution should be pushed as much as possible while sustaining adequate SNR throughout the brain (i.e. SNR > 10). Nonetheless, further valuable lessons can be translated from extensive YA-HCP trials which emphasized the importance of acquiring a large number of diffusion encoding directions (i.e. 270 directions or more) to mitigate the challenges in identifying crossing fiber bundles (Glasser et al., 2016b; Sotiropoulos et al., 2013). For example, isotropic resolution of 0.9 mm and 1.2 mm in macaque and human, respectively, yields approximately 60% of voxels containing three crossing fiber bundles (threshold at 0.05 of third crossing fiber's volume fraction) (Fig. 10a) (Autio et al., 2020; Sotiropoulos et al., 2013). Since the detection of crossing fiber architecture critically depends on CNR, we advocate a standard set of b-values (0, 1000, 2000 and 3000 s/mm$^2$) used by several recent large-scale human MRI consortia. Importantly, phase-reversed SE echo-planar images should be acquired to correct for geometric distortion (Fig. 3). We recommend utilizing monopolar gradients to minimize TE and maximize SNR. Moreover, multiband imaging substantially improves the efficiency of diffusion acquisition up to a point that incomplete T1-relaxation overrides the gains (i.e. TR ~ 3.0 sec). Clearly, efficient isotropic high-resolution dMRI acquisition in vivo requires high-density coils with good parallel imaging capabilities, whereas ex vivo studies can alleviate this requirement by using gadolinium contrast agents to shorten T1-relaxation constants and using longer data acquisition sessions. Finally, we recommend using prescan normalization to improve quality of image registration (dMRI to structural) and to reduce the effects of head motion within a static receive field.

To explore the NHP dMRI data quality, we also analyzed several PRIME-DE datasets using the HCP-NHP pipeline and FSL's 'bedpostx_gpu' (Autio et al., 2020; Behrens et al., 2007; Hernandez-Fernandez et al., 2018). Whole brain SNR of b0 images are compared in Figure 10a. Notably, the PRIME-DE data (MtS-P and UC-Davis) resulted in much lower third crossing fiber white matter volume fractions in comparison to RIKEN data despite using comparable imaging resolutions (RIKEN 0.7 mm$^3$, MtS-P 1.0 mm$^3$ and UC-Davis 0.7 mm$^3$) (Fig. 10b). This discrepancy may reflect the differences in number of diffusion-weighting gradient directions and number of b-values, which are important at these modest resolutions with respect to the size of white matter bundles.

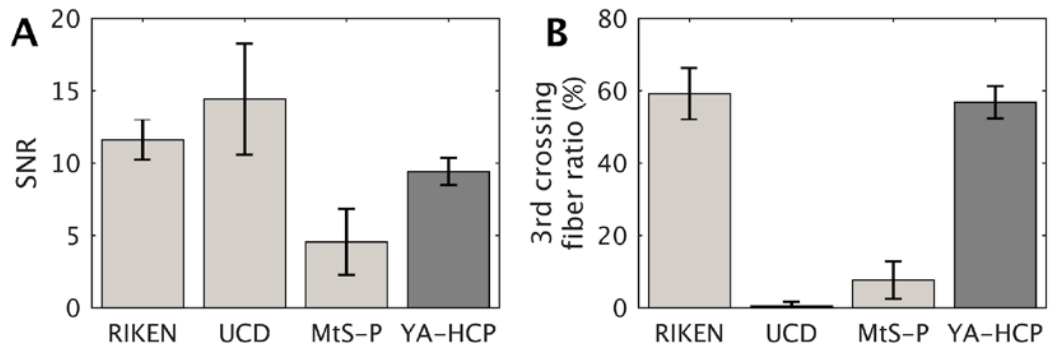

**Figure 10. Comparison of dMRI quality measures between imaging centers. (A)** Whole brain signal-to-noise ratio (SNR). **(B)** Third crossing fiber ratio in white matter (threshold at volume fraction 0.05). Primary imaging parameters were: RIKEN (N=20), MtS-P (N=4), UC-Davis (N=8) and YA-HCP (N=20): voxel size 0.7, 1.0, 0.7 and 2 mm$^3$; number of directions (512, 60, 121 and 256) and b-values (0, 1000, 2000 and 3000; 0 and 1000; 0 and 1600; 0, 1000, 2000 and 3000 s/mm$^2$), respectively. RIKEN data was obtained using HCP-style image acquisition protocols. Abbreviations: UC-Davis University of California, Davis; MtS-P Mount Sinai-Philips; YA-HCP the Young Adult Human Connectome Project.

Ex vivo studies provide the gold-standard for quantitative neuroanatomical connections and also the upper limit for tractography. A recent ex vivo marmoset study achieved a ~2000-fold smaller voxel size (80 µm isotropic) in comparison to YA-HCP's 7T dMRI (1 mm isotropic), providing details of primate neuroanatomy with unprecedented accuracy (Liu et al., 2020). Such efforts in combination with quantitative track tracing and histology are expected to yield important insights to establish more robust dMRI specifications.

3. *Discussion*

We have proposed a set of minimal specifications aimed at advancing NHP neuroimaging acquisition and analysis (Box 1). We have also demonstrated compelling benefits of cutting-edge accelerated imaging in NHP MRI, highlighting the advantages provided by NHP dedicated multi-channel RF coil technologies. Since most of the recommended minimal acquisition specifications have been tested and used by many recent large-scale human neuroimaging initiatives (i.e. HCP, UK BioBank, ABCD, Brain/MINDS), these guidelines should help improve NHP study reproducibility and help bridge the gap between NHP and human neuroscience. Although our primary focus was centered on issues surrounding NHP MRI, the guidelines are also applicable to other species such as rodents.

3.1. Towards reproducible NHP neuroimaging

NHPs are critical research models in basic and translational science due to their evolutionary proximity and similarity to humans in genetics, brain connectivity and behavior (Harding, 2017). However, systematic issues surrounding robustness, low statistical power due to a small sample size and lack of standardization result in translational studies in general having typically low study reproducibility (<25%) in preclinical biomedicine, pharmacology and neurosciences (Baker, 2016; Begley and Ellis, 2012; Glenn and Ioannidis, 2015; Prinz et al., 2011). Conversely, particularly low study reproducibility in human neuroimaging (Poldrack et al., 2017) may, in turn, adversely influence preclinical study designs. To establish a modern, more efficient, basic research and translational platform, it is imperative to investigate reproducibility in preclinical and human neuroimaging and to adapt best available practices (Glasser et al., 2016b; Nichols et al., 2017; Poldrack et al., 2017).

To improve the prospects for NHP neuroimaging, we introduced minimal NHP MRI standards (Box 1). Importantly, we demonstrated that in macaque monkeys these guidelines can enable automated and robust estimation of cortical thickness that were equivalent or higher than those reported in human subjects (Fig. 11a, b, Supp. Fig. 4, 5, 6). On the other hand, despite three decades of fMRI studies, rfMRI reproducibility remains a formidable challenge for NHP studies (Fig. 8b, Fig. 11d). Surprisingly, results show that pooling current rfMRI data across NHP imaging centers in PRIME-DE reduces rather than increases statistical strength, arguing against the primary aims of data sharing and reinforcing the challenges surrounding reproducibility in preclinical sciences. The low reproducibility may be heavily impacted by differences in anesthesia and data acquisition protocols (e.g. scanner, coil and sequence) used in different research centers. However, as more NHP imaging centers refine their experimental methodology according to more standardized image acquisition approaches (i.e. Box 1), in conjunction with improved anesthesia protocols and more stringent quality control (Vanduffel, 2018), we remain optimistic for pooling multi-center NHP functional data in the future.

Although we anticipate that most of the minimal NHP MRI guidelines are uncontroversial (Box 1), the main bottlenecks to adapt these specifications are related to hardware limitations (i.e. field strength, availability of gradient inserts, RF multi-array receive-coil and parallel imaging sequences) and the lack of automated analysis pipelines optimized for NHPs. Some of the hardware limitations can be mitigated as multiple imaging centers already have developed NHP dedicated multi-channel receive coil arrays (Autio et al., 2020; Ekstrom et al., 2008; Gao et al., 2020; Gilbert et al., 2016; Janssens et al., 2013, 2012; Schaeffer et al., 2019; Yacoub et al., 2020) that provide very cost-effective means to improve the NHP MRI data quality (Box 2, Fig. 4). Multichannel receive coil arrays are becoming commercially and thus more widely available for awake and anesthetized macaque (24-channel) (Autio et al., 2020) and marmoset (16-channels) monkeys for Siemens 3T and 7T scanners (https://www.rogue-research.com/takashima-seisakusho-coils/) and (implantable) arrays can also be made available through the Vanduffel lab.

Second, automatic pre-processing software, such as the HCP-NHP pipeline, is becoming available, but its *implementation* remains challenging. One way to overcome this is to establish an on-line system for automated pre-processing, similar to ones emerging in the human neuroimaging community (Esteban et al., 2020). As for statistical analyses, we advocate toolboxes firmly established for human neuroimaging data such as FSL's PALM (Permutation Analysis of Linear Models) (Winkler et al., 2014) which can operate over several modes of data (i.e. volume, CIFTI, GIFTI MRI data but also non-imaging data) and are rigorously validated using simulated and real data for controlling multiple comparisons (for best statistical practices, see (Eklund et al., 2016; Nichols et al., 2017)). We also advocate data sharing (e.g. raw and pre-processed) in public repositories, such as PRIME-DE (Milham et al., 2018), and published figures in the HCP's Connectome Workbench 'scene' file format at Brain Analysis Library of Spatial maps and Atlases (BALSA) (Van Essen et al., 2017) to improve comparison across studies, and new initiatives of the Human Brain Project. Altogether, technology and data sharing platforms are available for the NHP community to substantially improve robustness and reproducibility in NHP neuroimaging.

Ultimately, NHP MRI guidelines (Box 1) should be validated and improved upon with respect to the anatomical and physiological 'ground truth'. NHP research has the potential to provide important insights for improving human neuroimaging (Orban et al., 2004; Vanduffel et al., 2014), as imaging data and preprocessing strategies can be compared in the same subjects with electrophysiology (Tsao et al., 2006) and post-mortem data (Hayashi et al., this issue).

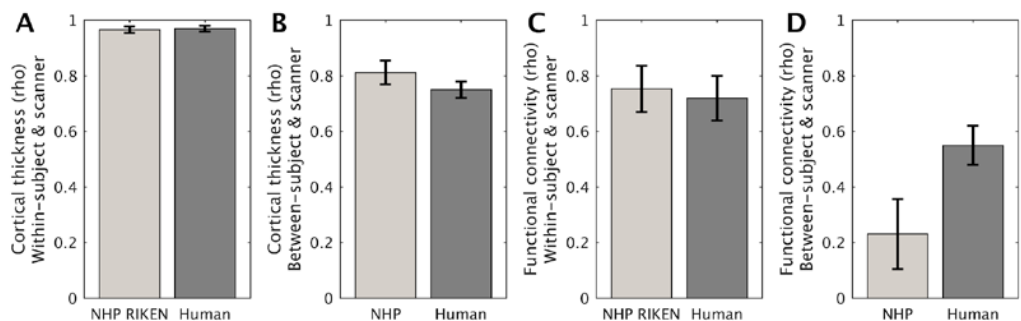

**Figure 11. Comparison between heterogeneous non-human primate (NHP) protocols and human harmonized protocol (HARP) MRI similarity measures.** Parcellated cortical thickness similarity measures were comparable across species **(A)** within-subject and within scanner (at RIKEN, N=5) and **(B)** between-subject and between-scanners (macaque N=23, 6-scanners; human N=30, 13-scanners). **(C)** Parcellated functional connectivity (FC) exhibited comparable test-retest reproducibility between anesthetized NHPs in RIKEN and (awake) humans. However, **(D)** FC exhibited poor reproducibility across NHP imaging centers in comparison to humans. HARP is a travelling subject (N=30) study across 13-clinical MRI centers (Koike et al., 2020). Macaque and human data were processed using HCP-NHP and HCP pipelines, respectively. NHP MRI data was parcellated using M132 91-areas per hemisphere atlas (Markov et al., 2014) whereas human MRI data was parcellated using HCP's 180-areas per hemisphere atlas (Glasser et al., 2016a). The Spearman's rank correlation coefficients (rho) are shown in mean (std).

3.2. Challenges in multi-center NHP research

Multi-center approaches would enable investigations of much larger numbers of NHP subjects with a more multidisciplinary approach than possible in a single laboratory. However, substantial challenges remain to remove laboratory-specific non-biological measurement biases from multi-center NHP MRI data using prospective data acquisition and retrospective data analysis harmonization methodologies. In humans, prospectively harmonized multi-center data acquisition protocols and ICA-FIX cleaned data result in FC matrices across subjects (N=30) and scanners (n=13) is 0.55 ± 0.07 (rho ± std) whereas within-subject and within-scanner test-retest reproducibility is 0.72 ± 0.08 (Fig. 10d) (Koike et al., 2020). Thus, the average FC correspondence between NHP imaging centers (0.23 ± 0.13; rho ± std) is well below that observed across human MRI centers (which use much more convergent hardware and harmonized data acquisition protocols) (Fig. 10d), whereas moderate within-subject and within-scanner test-retest reproducibility correspondence within RIKEN (0.75 ± 0.08; rho ± std) show promising directions for systematically increasing reproducibility (Fig. 8, Fig. 10c).

Removal of undesirable non-biological sources of variation in multi-center studies is a rapidly emerging area of investigation and different retrospective harmonization strategies have been introduced for cortical thickness (Fortin et al., 2018), FC (Feis et al., 2015; Yu et al., 2018) and diffusion tensor imaging (Fortin et al., 2017). Nonetheless, a consensus has not been achieved over optimal harmonization methods (i.e. General Linear Model, ICA, empirical Bayes and convolutional neural network) to selectively remove site-related effects while maintaining biologically relevant covariates in the data. Statistical harmonization in the NHP population, however, remains a daunting challenge given that the majority of the underlying biological co-factors (e.g. phenotyping data) remain unknown. For perspective, a recent study of UK Biobank study found significant associations between brain volumetric phenotypes and over 100 associated genes, but relied on a very large population (~20,000 individuals) (Zhao et al., 2019). Such populations are unrealistic for NHP studies. To overcome this limitation, NHP community should adapt best possible MRI practices.

4. Conclusion

Here, we provided minimal acquisition guidelines for NHP MRI with an aim to standardize data acquisition and data analysis with respect to the human neuroimaging community. Increased standardization will not only serve the needs of human neurosciences and clinical services but will also improve prospects of translating clinical findings to basic and pre-clinical NHP neuroimaging. However, much remains to be done in terms of multi-center NHP MRI research and the way forward is by encouraging greater dialogue and cooperation including sharing data acquisition technologies, image processing software, and data access now facilitated via emerging collaborative initiatives such as PRIME-DE.

**Notes**

Supplementary Information is available in the online version of the paper.

**Acknowledgements**

We express gratitude to the Primate Neuroimaging Data-Exchange (PRIME-DE) initiative and to all institutions that contributed to the PRIME-DE dataset. Human data were provided by the Human Connectome Project, WU-Minn Consortium (Principal Investigators: David Van Essen and Kamil Ugurbil; 1U54MH091657) funded by the 16 NIH Institutes and Centers that support the NIH Blueprint for Neuroscience Research; and by the McDonnell Center for Systems Neuroscience at Washington University. This research is partially supported by the program for Brain/MINDS and Brain/MINDS-beyond from Japan Agency for Medical Research and development, AMED (JP20dm037006, T.H.), by RIKEN Compass to Healthy Life Research Complex Program from Japan Science and Technology Agency, JST, by JSPS KAKENHI Grant Number JP20K15945 (J.A.A.), by the European Research Council (ERC) under the European Union's Horizon 2020 research and innovation programme (Grant agreement No. 802482, C.M.S.), the Canadian Institutes of Health Research FRN-148453 and Natural Sciences and Engineering Research Council of Canada Discovery Grant (R.S.M), EB027061 (J.Z.), DA048742 (J.Z.), NSF-2024581 (J.Z.), UMN Digital Technologies Initiative (J.Z.), UMN AIRP (J.Z.), Lynne and Andrew Redleaf Foundation (D.A.F.), DA041148 (D.A.F.), DA04112 (D.A.F.), MH115357 (D.A.F.), MH096773 (D.A.F), RF1 MH116978 (E.Y.), P50NS098573 (E.Y.), the Emmy Noether Program of the German Research Foundation (SCHW1683/2-1, C.M.S), NIH Grant MH-060974 (D.C.V.E.) NIH Grant BRAINS R01-MH101555 and RF1MH117040 (B.E.R) and KU Leuven grant C14/17/109; Fonds Wetenschappelijk Onderzoek-Vlaanderen (FWO) G0D5817N, G0B8617N, G0C1920N, G0E0520N, VS02219N; and the European Union's Horizon 2020 Framework Programme for Research and Innovation under Grant Agreement No 945539 (Human Brain Project SGA3) (W.V). The authors have no conflicts of interest to declare.

**Author contributions**
Joonas A. Autio: Conceptualization, Investigation, Formal Analysis, Writing - original draft, review & editing.
Qi Zhu: Investigation, Formal Analysis, Review & editing.
Xiaolian Li: Investigation, Formal Analysis, Review & editing.
Matthew F. Glasser: Conceptualization, Review & editing.
Caspar M. Schwiedrzik: Review & editing.
Essa Yacoub: Funding acquisition, Review & editing.
Damien A. Fair: Review & editing.
Jan Zimmermann: Review & editing.
Ravi S. Menon: Review & editing.
David C. Van Essen: Conceptualization, Funding acquisition, Review & editing.
Takuya Hayashi: Conceptualization, Funding acquisition, Investigation, Formal Analysis, Review & editing.
Brian Russ: Writing - original draft, Review & editing.
Wim Vanduffel: Conceptualization, Funding acquisition, Investigation, Formal Analysis, Writing - original draft, Review & editing.